# Comprehensive Investigation of Fundamental Mode Profiles in Monolithic Nonplanar Ring Oscillators

**WEITONG FAN[†], CHUNZHAO MA[†], WENXUN LI, DANQING LIU, ZELONG HUANG, JIE XU, HSIEN-CHI YEH, AND CHANGLEI GUO***

*MOE Key Laboratory of TianQin Mission, TianQin Research Center for Gravitational Physics & School of Physics and Astronomy, Frontiers Science Center for TianQin, CNSA Research Center for Gravitational Waves, Sun Yat-sen University (Zhuhai Campus), Zhuhai 519082, China*

[†]*These authors contributed equally to this work.*

** guochlei@mail.sysu.edu.cn*

**Abstract:** Nonplanar ring oscillators (NPROs) are building blocks for high-performance single-frequency lasers and ring-laser gyroscopes that have profoundly improved the state-of-the-art laser technologies, fundamental research and precision measurements. However, a comprehensive investigation of fundamental mode profiles in monolithic NPROs has been a missing part even though they will affect the performance of the lasers or ring-laser gyroscopes. Here, we present a comprehensive finite-element modeling of the output beam profiles of monolithic NPROs by combining ABCD transmission matrix and generalized Huygens-Fresnel integral. We theoretically investigate the effects of geometric parameters of monolithic NPROs on their output mode profiles. In particular, we focus on the thermal effect inside the monolithic NPRO and calculate the equivalent focal-length of the thermal-lens by using a ray-tracing-method. Furthermore, we experimentally characterize the output laser beam profile, reconstruct the beam profile at the *A*-facet of the monolithic NPRO, and compare the experimental results with the simulation, thereby validating the accuracy and reliability of our model. The investigation may facilitate monolithic NPRO design and subsequently improve the performances of NPRO lasers and gyroscopes in the future.

## 1. Introduction

The concept of nonplanar ring oscillators (NPROs) based on Nd:YAG was first proposed by A.R. Clobes in 1972 [1]. Exploiting the Sagnac effect, the NPRO laser has also been successfully employed as ring-laser gyroscopes for rotational sensing [2–5]. With the continuous advancement of laser technology, the monolithic nonplanar ring oscillator was reported by Kane in 1985 [6]. This innovative design integrates the gain medium and the resonator into a single unit [7], establishing itself as a high-performance laser that has been widely utilized in industrial, scientific, and space missions to the present day [8–14]. Recent report has validated the application of monolithic NPRO in laser gyroscope, demonstrating the significant potential of this type of NPRO for use in gyroscopic applications [15]. In both fields of single-frequency lasers and ring-laser gyroscopes, the shapes of output fundamental mode profiles from NPRO lasers are of utmost importance, because a pure Gaussian beam is preferred for many applications. In precision measurements like gravitational wave detection, the interference highly relies on the injection of the fundamental transverse mode into the interferometer [16–18]. Thus, ensuring the mode purity of the laser output and the beam quality of the fundamental mode are crucial. Moreover, the mode profiles are also strongly correlated with thermal effects in the resonator, where the latter is crucial for laser design especially for optical power scaling [19–24]. Therefore, a comprehensive investigation of fundamental mode profiles in monolithic NPROs is very necessary.

In previous studies on NPROs [25,26], methods have been established for calculating the output beam profiles. However, these methods are primarily applicable under the conditions of large resonator using discrete optics, as well as in equilateral ring configurations (pertain to large-scale laser gyroscopes). Such approaches are not suitable for smaller, non-equilateral monolithic NPROs. It is noteworthy that existing studies [27,28] on monolithic NPRO still suffer from critical limitations: the lack of detailed derivations for key parameters undermines model transparency and reproducibility, while the limited resolution impedes precise characterization of beam profiles. These issues severely hinder both theoretical guidance and practical optimization for high-precision NPRO design.

Furthermore, the thermal field induced by pump lasers can significantly affect the output beam profiles, yet this aspect has not been thoroughly addressed in previous studies [28–32]. Although the thermal-lens effects in monolithic NPROs have been studied [33], the focal lengths obtained through integral calculations remain insufficiently intuitive, and there has been no experimental validation to support their findings. To date, there has been still a lack of reports addressing the output beam profiles of widely utilized monolithic NPRO, particularly studies that encompass various sizes and design configurations, as well as the alterations in the beam profiles induced by thermal-lensing effects at different pump levels. Thus, a gap remains between a complete knowledge of monolithic NPRO mode profile and its more and more profound applications in high-performance laser technologies.

In this work, we present a comprehensive study on the output beam profiles of monolithic NPROs with a more flexible calculation method. We systematically investigate the effects of various geometrical parameters, including the out-of-plane angle, incident and output laser angles, and the dimensions, on the spatial distribution of the output beam profiles using finite-element method (FEM). Furthermore, we place significant emphasis on the effects of thermal-lens within the monolithic NPRO. We calculate the equivalent focal length of the thermal-lens through physical field simulations and ray-tracing-method, presenting a more intuitive physical result and analyze the impact of thermal-lens on the output beam profile, which is further validated by experimental results. The results of this work can optimize the design of monolithic NPROs used for single-frequency lasers and gyroscopes [15,34–36].

## 2. ABCD Matrix of beam transmission in monolithic NPRO

Figure 1(a) illustrates the schematic structure of a typical monolithic NPRO, which consists of one output coupling facet and three total internal reflection (TIR) facets. Point *A* is located at front facet, which is coated for anti-reflection at pump wavelength and high-reflection at laser wavelength. The facets containing points *B*, *C*, and *D* are optically polished planes where total internal reflection occurs. The geometric structure of the light path within the resonator is shown in Fig. 1(b), defined by four non-coplanar reflective facets, where $\beta$ is the dihedral angle between the two transmission planes. $\theta_A$ and $\theta_B$ are half the angle between the incident and reflected beams. Point *E* is the midpoint of the *BD* connection. We assume that the beam within the resonator originates from point ***S*** which is on leg *AB* and infinitely close to point *A*, propagating in a clockwise direction along the *A*-*B*-*C*-*D*-*A* path.

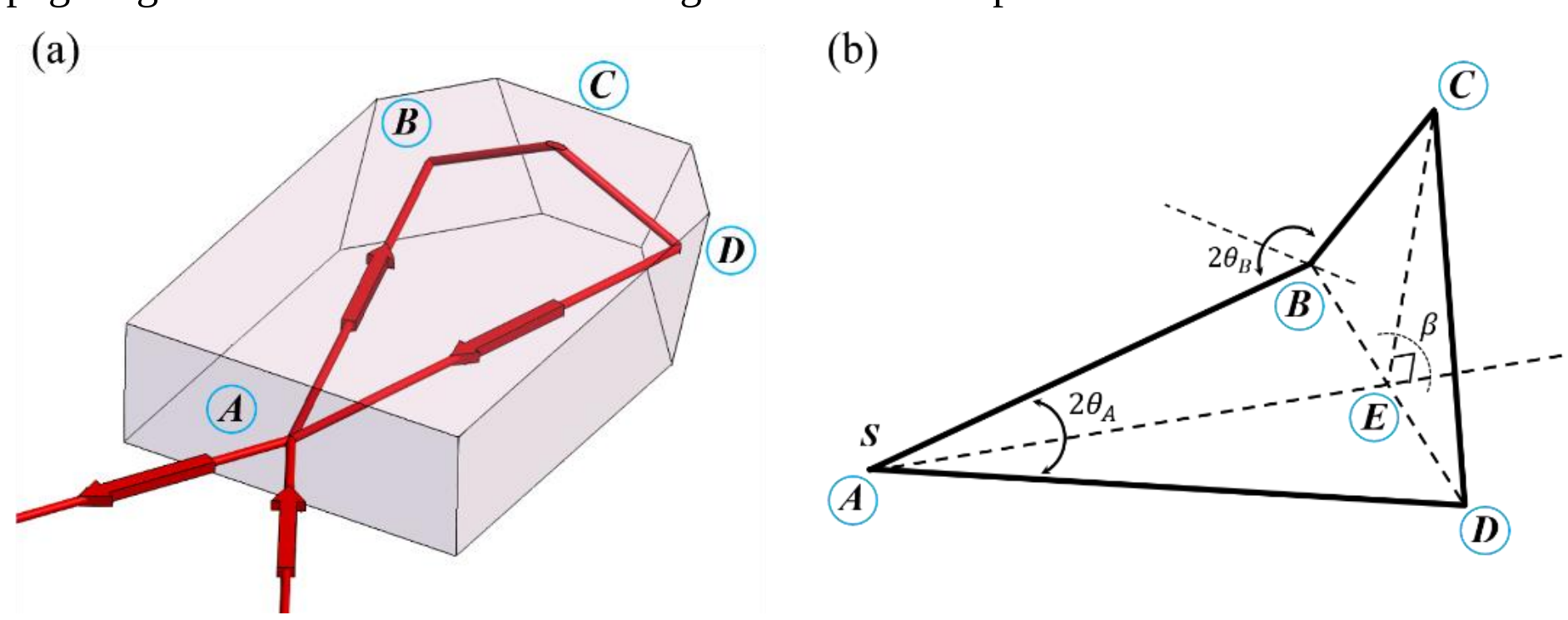

Fig. 1. (a) The structure and optical path of monolithic NPRO. Red traces mark the optical path along *A-B-C-D-A*. (b) The geometric structure of the light path in monolithic NPRO. *AE*, *CE* and *BD* are auxiliary lines for the nonplanar ring, and *E* is the midpoint of *BD*. $\theta_A$, $\theta_B$ and $\beta$ are also defined here.

Figure 2 demonstrates the coordinate transformations among different main axis systems. During one round trip within the resonator, the beam undergoes four reflections, and the nonplanar characteristics of the resonator result in the incident planes of successive reflections being non-coplanar. Therefore, it is necessary to describe both the incident and reflected waves in the main axis system, where the vector $\boldsymbol{z}$ aligns with the direction of beam propagation, vector $\boldsymbol{y}$ lies within the transmission plane and is orthogonal to $\boldsymbol{z}$, and vector $\boldsymbol{x}$ is determined by the right-hand rule. The beam propagating along *AB* reflects at point *B*, resulting in a rotational transformation of the vectors between the incident and reflected coordinate systems. The vectors $\boldsymbol{x}$, $\boldsymbol{y}$, $\boldsymbol{z}$ will first undergo reflection transformations, with $\mathrm{y}_{AB}$ transitioning to $\mathrm{y}'_{AB}$ , while the vector $\boldsymbol{z}$ continues along the direction of the reflected beam.

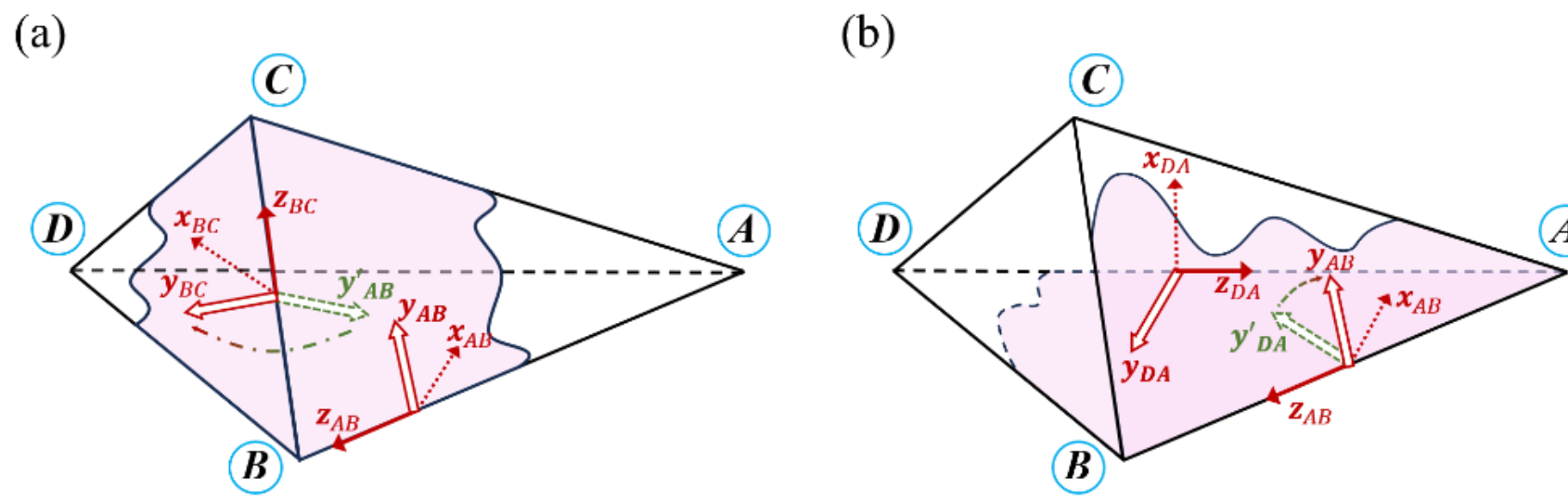

Fig. 2. The coordinate transformations among different main axis systems. (a) From leg *AB* to *BC*; (b) From leg *DA* to *AB*. The $\boldsymbol{x}$, $\boldsymbol{y}$, $\boldsymbol{z}$ vectors are used to define the direction of beam propagation and spatial coordinate changes. The pink area represents the plane where the light beam propagates.

To ensure that the description of the transmission loop strictly adheres to the new principal axis system, the $\boldsymbol{x}$, $\boldsymbol{y}$, $\boldsymbol{z}$ vectors will undergo another spatial coordinate rotation, maintaining the vector $\boldsymbol{z}$ along the direction of propagation, while the vector $\boldsymbol{y}$ adjusts to lie within the new transmission plane *BCD* ( $\mathrm{y}_{BC}$ ). The $\boldsymbol{x}$ vector is determined by the right-hand rule, perpendicular to the new transmission plane. Consequently, in the context of continuous mirror reflections involving different axis systems, it becomes necessary to introduce a rotation matrix that projects the vectors from the old coordinate system to the new one. The rotation matrix is expressed as Eq. (1).

$$M(\beta) = \begin{bmatrix} \cos(\beta) & -\sin(\beta) & 0 & 0 \\ \sin(\beta) & \cos(\beta) & 0 & 0 \\ 0 & 0 & \cos(\beta) & \sin(\beta) \\ 0 & 0 & -\sin(\beta) & \cos(\beta) \end{bmatrix} \tag{1}$$

In conventional resonator theory, paraxial beam would escape from cavities formed by four planar mirrors, preventing resonance. The NPRO, due to the existence of internal thermal-lens, constrain the paraxial beam and resonate. We equivalently represent the thermal-lens introduced by temperature-induced refractive index gradients in the beam loop as two concave lenses located at points *A* and *B* (detailed analysis will be conducted in Section. 4). Therefore, the reflection matrix *Mr* at points *A* and *B* can be replaced with a spherical mirror transformation matrix, as shown in Eq. (2), and *f* is the equivalent focal length of the thermal-lens. The placement of the spherical lens at point *B* is primarily intended to enhance the model's accuracy under high pump-power.

$$M(\theta)=\begin{bmatrix} 1 & 0 & 0 & 0 \\ 0 & 1 & 0 & 0 \\ -1/f/\cos(\theta) & 0 & 1 & 0 \\ 0 & -\cos(\theta)/f & 0 & 1 \end{bmatrix} \tag{2}$$

The free propagation matrix of the beam within the medium can be expressed as Eq. (3), where *L* is the length of the laser propagation and *n* is the refractive index of the gain medium.

$$M(L)=\begin{bmatrix} 1 & 0 & L/n & 0 \\ 0 & 1 & 0 & L/n \\ 0 & 0 & 1 & 0 \\ 0 & 0 & 0 & 1 \end{bmatrix} \tag{3}$$

Utilizing the ABCD matrix for beam propagation, we can derive the expression for the beam transmission loop after one round trip starting from point ***S*** within the resonator, as presented in Eq. (4).

$$\begin{aligned} M_{round} &= M\left(\beta_{D-AB-C}\right)M\left(\theta_A\right)M\left(L_{AD}\right)M\left(\beta_{C-AD-B}\right)M_r M\left(L_{CD}\right) \\ &\quad M\left(\beta_{B-CD-A}\right)M_r M\left(L_{BC}\right)M\left(\beta_{A-BC-D}\right)M\left(\theta_B\right)M\left(L_{AB}\right) \\ &= \begin{bmatrix} \mathrm{A}_{2\times2} & \mathrm{B}_{2\times2} \\ \mathrm{C}_{2\times2} & \mathrm{D}_{2\times2} \end{bmatrix} \end{aligned} \tag{4}$$

### 3. Transverse Profiles in monolithic NPRO

Combining the ABCD transmission matrix, the initial electric field $E_1$ originating from ***S*** and the profile $E_2$ after one-round-trip transmission can be theoretically studied by using the generalized Huygens-Fresnel integral, as shown in Eq. (5). Here, $r_i^T=\left(x_i, y_i\right), i=1,2$ , represents the transverse positions, where $i$ denotes the respective fields.

$$\begin{aligned} E_2\left(x_2, y_2, z_{\mathrm{cycle}}\right) &= \left(-\frac{i/\lambda}{|\mathrm{B}|^{1/2}}\right) exp\left(ikz\right)\iint dx_1 dy_1 E_1\left(x_1, y_1, 0\right) \\ &\quad \cdot exp\left\{\frac{ik}{2}\begin{bmatrix} r_1 \\ r_2 \end{bmatrix}^T \begin{bmatrix} \mathrm{B}^{-1}\mathrm{A} & -\mathrm{B}^{-1} \\ -\mathrm{B}^{-1} & \mathrm{D}\mathrm{B}^{-1} \end{bmatrix}\begin{bmatrix} r_1 \\ r_2 \end{bmatrix}\right\} \end{aligned} \tag{5}$$

According to the self-reproducing mode principle, certain optical beam within a resonator, after propagating through the cavity and returning to their initial position, maintain the same spatial distribution of optical field as when it originates. This leads to the following expression (6), where $\gamma$ represents the amplitude decay and phase change.

$$E_2\left(x_2, y_2, z\right)=\gamma E_1\left(x_1, y_1, 0\right) \tag{6}$$

Since it is not feasible to obtain an analytical solution by directly solving Eqs. (5) and (6), we resort to the FEM for analysis. We discretize the four facets of the monolithic NPRO, as illustrated in Fig. 3. A grid density of $g$=$33^2$ achieves acceptable theoretical error bounds while avoiding excessive computational load caused by overly fine mesh discretization. During the meshing process, we observe that each element on one mirror interacts with every element on the subsequent mirror, and similarly, all elements on one mirror will influence each element on the next mirror [37]. Consequently, the electric field $E_1$ can be represented by $E_1=\left[\mathrm{e}_1, \mathrm{e}_2, \mathrm{e}_3, \cdots\cdots, \mathrm{e}_\mathrm{g}\right]^T$ , where g represents the total number of discretized elements. Additionally, each small element $\mathrm{e}_j$ corresponds to different position, represented by $r_{ij}^{\ T}=\left(x_{ij}, y_{ij}\right)$ with $i=1,2$ and $j=1\sim g$ .

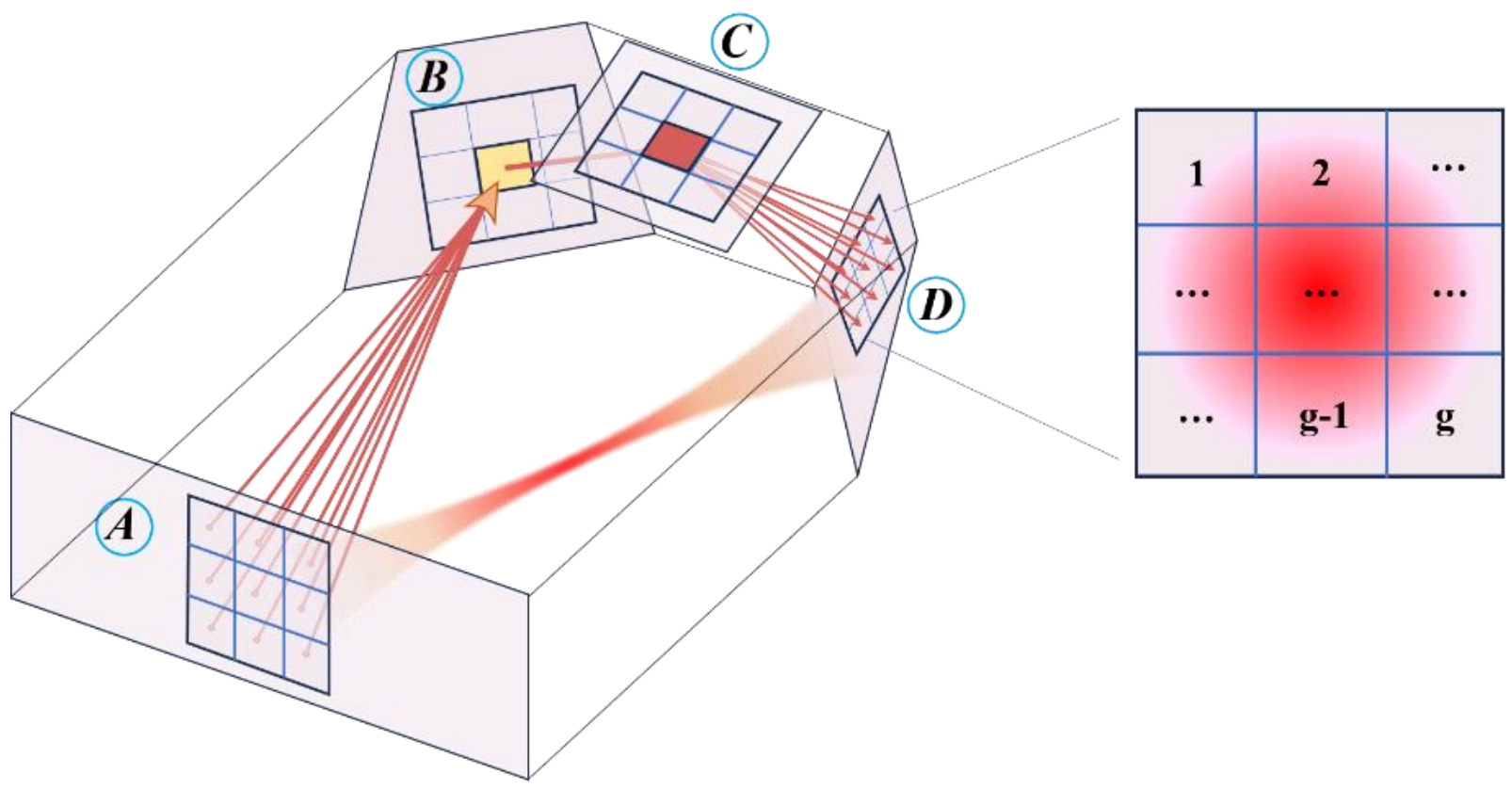

Fig. 3. Discretization of the four facets of monolithic NPRO. In subsequent calculations, each facet with dimensions of $330\,\mu\text{m}\times 330\,\mu\text{m}$ will be divided into $33^2$ smaller units in a square configuration ($g=33^2$).

The amplitude within each element can be considered as invariant when the number of discretized elements is sufficiently large, thus allowing us to express $E_1$ in Eq. (5) independent of the variations in the integration variables *x* and *y*. Therefore, we can rewrite Eq. (5) as the following Eq. (8). The eigenvectors of matrix *V* characterize the complex electric field at the position *S*, while the eigenvalues represent the corresponding mode losses.

$$\gamma\begin{bmatrix} e_1 \\ e_2 \\ \vdots \\ e_g \end{bmatrix} = V\times\begin{bmatrix} e_1 \\ e_2 \\ \vdots \\ e_g \end{bmatrix},\quad V=\begin{bmatrix} v_{11} & v_{12} & \cdots & \cdots & v_{1g} \\ v_{21} & \ddots & & & \vdots \\ \vdots & & v_{jj'} & \ddots & \vdots \\ v_{g1} & \cdots & \cdots & \cdots & v_{gg} \end{bmatrix},\quad (j, j'=1\sim g) \tag{7}$$

$$v_{jj'}=\left(-\frac{i/\lambda}{|\mathrm{B}|^{1/2}}\right)exp(ikz)\iint dx_1dy_1 \cdot exp\left\{\frac{ik}{2}\begin{bmatrix} r_{1j} \\ r_{2j'} \end{bmatrix}^T\begin{bmatrix} \mathrm{B}^{-1}\mathrm{A} & -\mathrm{B}^{-1} \\ -\mathrm{B}^{-1} & \mathrm{D}\mathrm{B}^{-1} \end{bmatrix}\begin{bmatrix} r_{1j} \\ r_{2j'} \end{bmatrix}\right\} \tag{8}$$

Typically, the geometrical parameters of NPROs are primarily controlled by four independent variables: $AE$, $CE$, $\theta_A$, and $\beta$, as shown in Fig. 4. $AE$ and $\theta_A$ jointly determine the length of optical path within the resonator, while $CE$ and the out-of-plane angle $\beta$ define the height of the resonator. In the subsequent calculations, we will use the monolithic Nd:YAG (0.8 at.%) NPRO ($3\text{mm}\times 8\text{mm}\times 12\text{mm}$) employed in experiments as a reference, specifically with parameters $AE=10.5\,\text{mm}$, $CE=1.5\,\text{mm}$, $\theta_A=15°$, $\beta=90°$ (with coating of 99.8% for S-polarization and 94% for P-polarization), to analyze the variations with different design parameters.

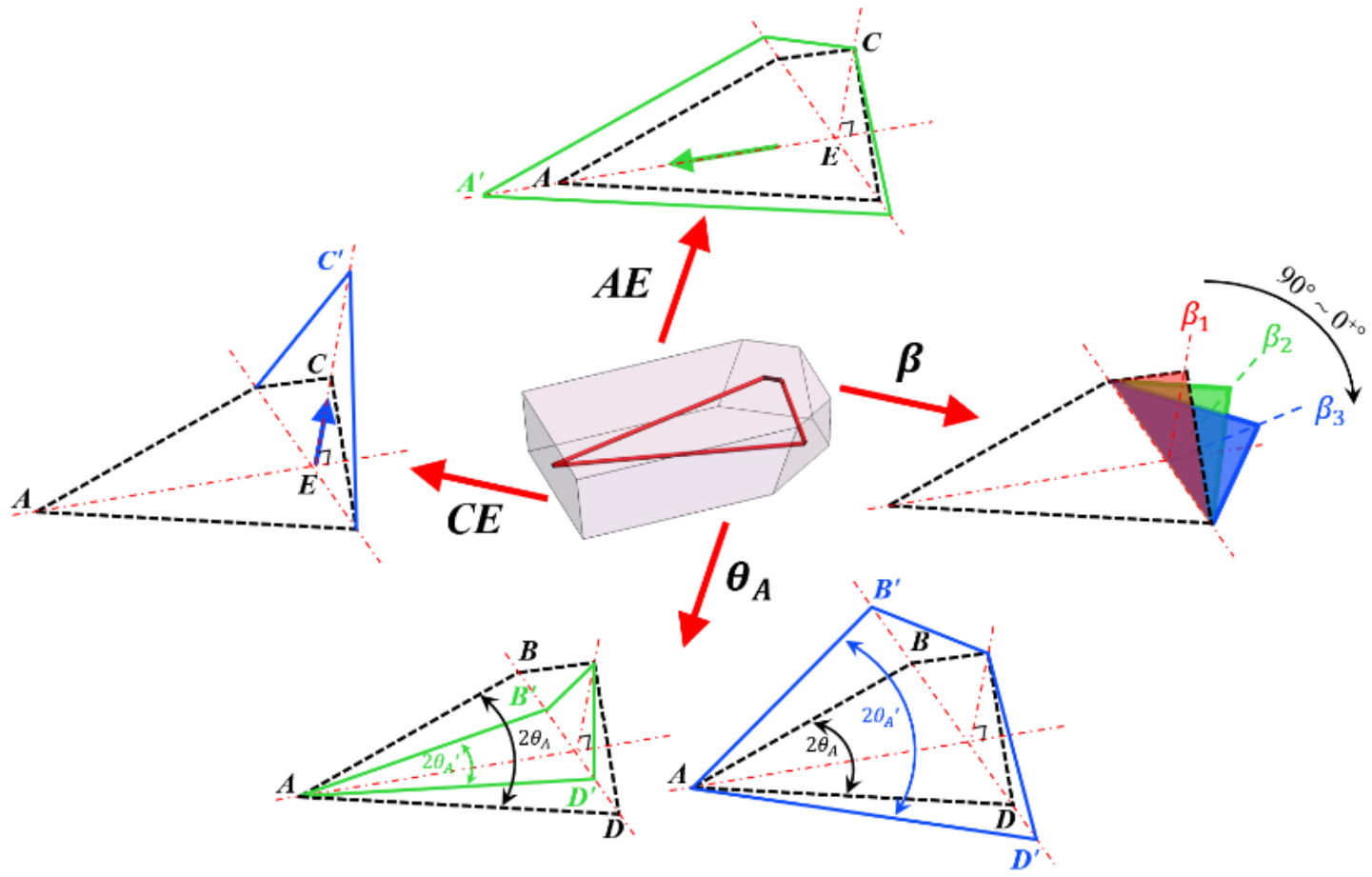

Fig. 4. Diagrams to show how the four critical parameters (*AE*, *CE*, $\beta$ and $\theta_A$) determine the geometry of monolithic NPROs.

By maintaining the length of *AE* or *CE*, while systematically varying the length of *CE* or *AE*, the beam profile at point ***S*** undergoes the changes as depicted in Fig. 5(a, b). To ensure total internal reflection, *AE*: CE must be greater than 3. The radius of the beam profile is shown by the red dashed line ($1/e^2$ of Gaussian fit). It can be observed that as *CE* progressively decreases, the beam radius correspondingly diminishes. Notably, when the ratio $AE:CE \geq 6$, the beam field's radius decreases slowly to a constant. This phenomenon occurs due to the fact that separation between the two concave lenses remaining constant. Contrary to the changing *CE*, the output beam profile does not exhibit a tendency towards a constant with the changing *AE*. This can be simply understood as the thermal-lens in the ring cavity being effectively located at points *A* and *B*. Changing the length of *AE* essentially modifies the distance between these lenses, thereby impacting the size of the output beam. The beam profile radius with changing *AE* (red curve) and *CE* (blue curve) are shown in Fig. 5(c).

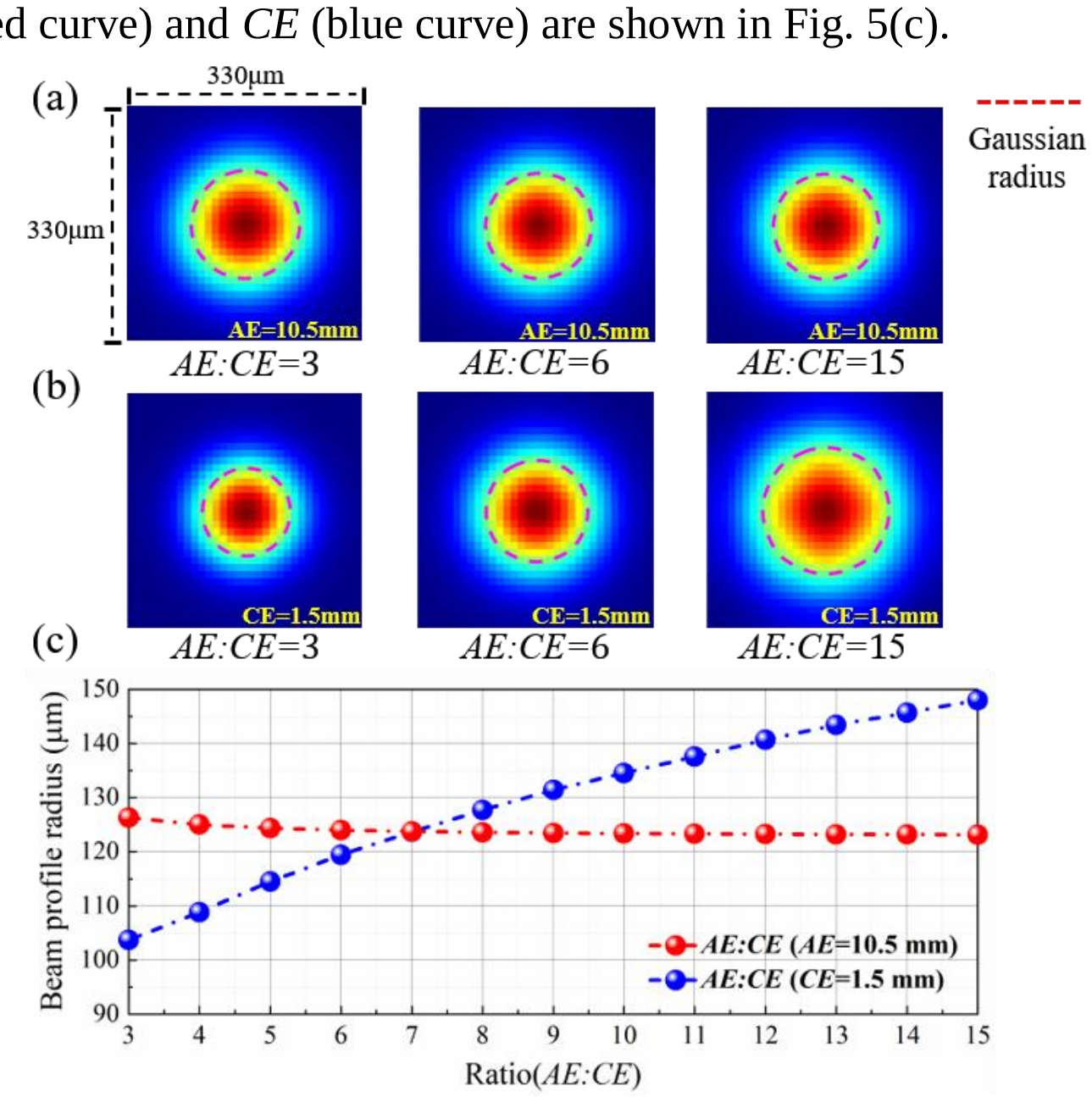

Fig.5. (a) The beam profiles at point ***S*** (see Fig. 1(b)) with different length of *CE*. (b) The beam profiles at point ***S*** with different length of *AE*. (c) Beam profile radius versus ratio of *AE* to *CE*. In the simulations, $\beta$=90°, $\theta_A$=15°, $f_A$=4 m, $f_B$=$10^5$ m

By varying the incident angle $\theta_A$ (with values selected near the selected near the total internal reflection boundaries of $0^{+\circ}$ to $24.5°$), the results are presented in Fig. 6(a). It can be observed that as the incident angle increases, the symmetry of the beam profile remains unchanged, indicating that the magnitude of the incident angle does not induce image rotation. This implies that, in crystal design, further miniaturization of the resonator can be achieved by reducing the incident angle $\theta_A$ to shorten the *BD* length.

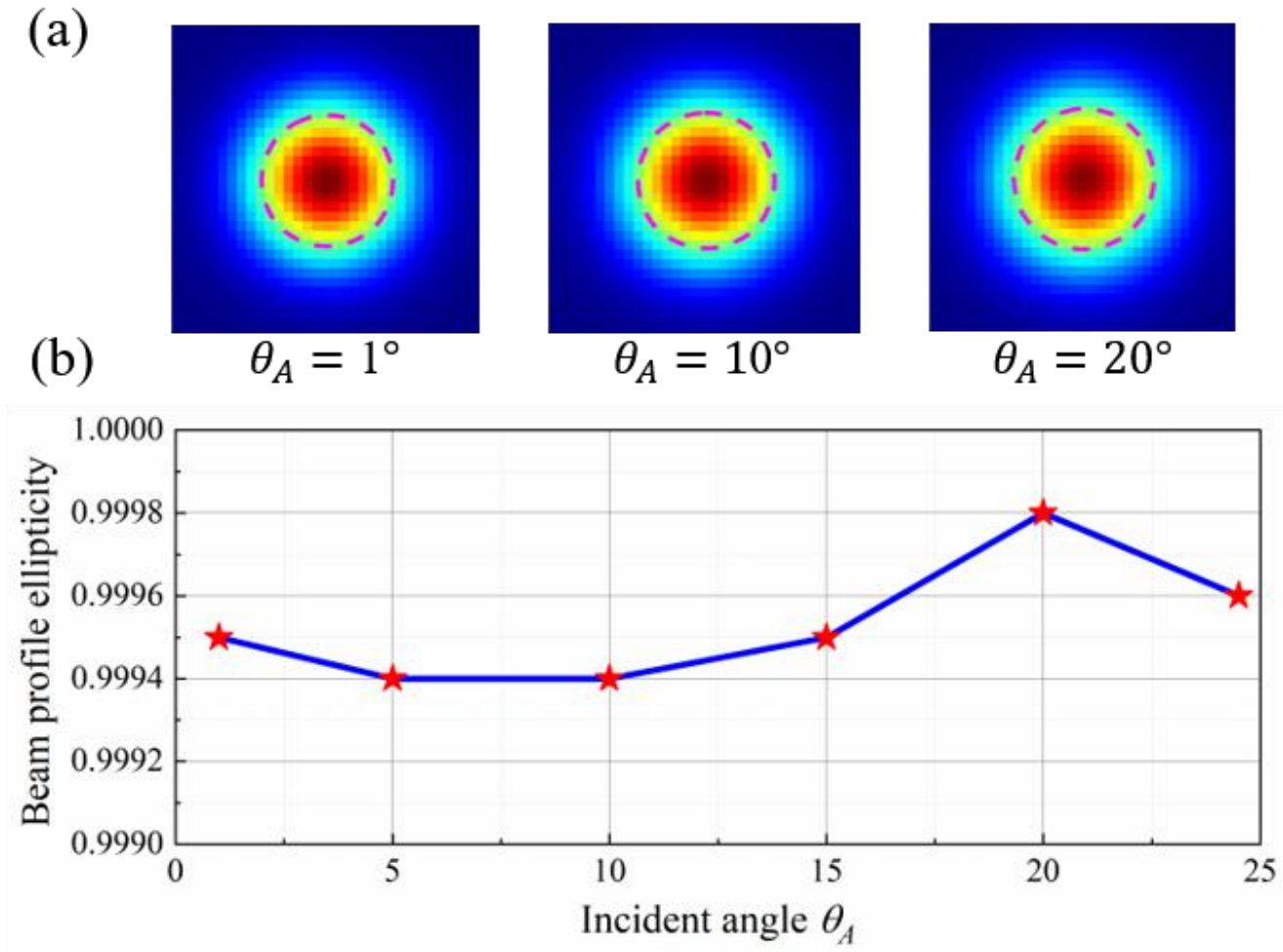

Fig. 6. (a) The beam profiles at point ***S*** with different incident angle $\theta_A$. (b) Beam profile ellipticity versus incident angle. In the simulations, *AE*=10.5 mm, *CE*=1.5 mm, $\beta$=90°, $f_A$=4 m, $f_B$=$10^5$ m.

In our theoretical calculations, we found that the out-of-plane angle of NPRO does not affect the output beam profile. However, previous studies have indicated that it influences the efficiency of magnetic field utilization along the beam path. Specifically, under the same magnetic field conditions, different out-of-plane angles result in varying loss differences between the clockwise and counterclockwise paths. The smaller is the out-of-plane angle, the lower is the required mode-selection magnetic-field strength.

## 4. Effects of Thermal-Lens on Beam Profile

In monolithic NPRO, thermal-lens effect arises primarily due to the absorption of pump laser by the gain medium within the cavity, which leads to localized heating (especially at the *A*-facet). This heating causes a change in the refractive index of the medium, creating a lens-like effect that can alter the path and focus of the laser beam. This phenomenon can significantly impact the performance and stability of the laser system, and has an influence on beam quality and alignment.

In the steady-state scenario, the internal temperature $T$ distribution of the monolithic NPRO can be derived by solving the heat conduction equation. Within the Cartesian coordinate framework, the heat conduction equation for the resonator is formulated as Eq. (9), with the coordinate system's origin set at the center of *A*-facet.

$$\frac{\partial^2 T}{\partial x^2}+\frac{\partial^2 T}{\partial y^2}+\frac{\partial^2 T}{\partial z^2}=-\frac{Q(x,y,z)}{K_c} \tag{9}$$

Here, $K_c$ represents thermal conductivity of the gain material, and $Q(x,y,z)$ denotes the heat source density within the resonator. The internal heat sources in the resonator arise from non-radiative transitions during operation. Distinct from direct heating, the conversion of external input $P_{abs}$ into thermal energy is attributed to the difference between the pump laser power $P_{pump}$ and the output laser power, as delineated in Eq. (10),

$$P_{abs} = (P_{pump} - P_{th}) \times \left(1 - \eta_{slope}\right) + P_{th} \tag{10}$$

where $P_{th}$ is the laser threshold power, and $\eta_{slope}$ is the laser slope efficiency. The heat source density can be expressed as Eq. (11),

$$Q(x, y, z) = \frac{2P_{abs}}{\pi\omega_p^2} \times \exp\left(\frac{-2\left(x^2 + y^2\right)}{\omega_p^2} - \alpha z\right) \tag{11}$$

where $\omega_p$ is the Gaussian radius of the pump light (assuming constant radius during propagation) and $\alpha$ is the absorption coefficient of the resonator.

The temperature distribution within the resonator under steady-state conditions was analyzed using FEM. Fig. 7 presents the thermal field within the resonator, temperature distribution of the axial section along leg *AB* and the *A*-facet, under a pump power of 1000 mW.

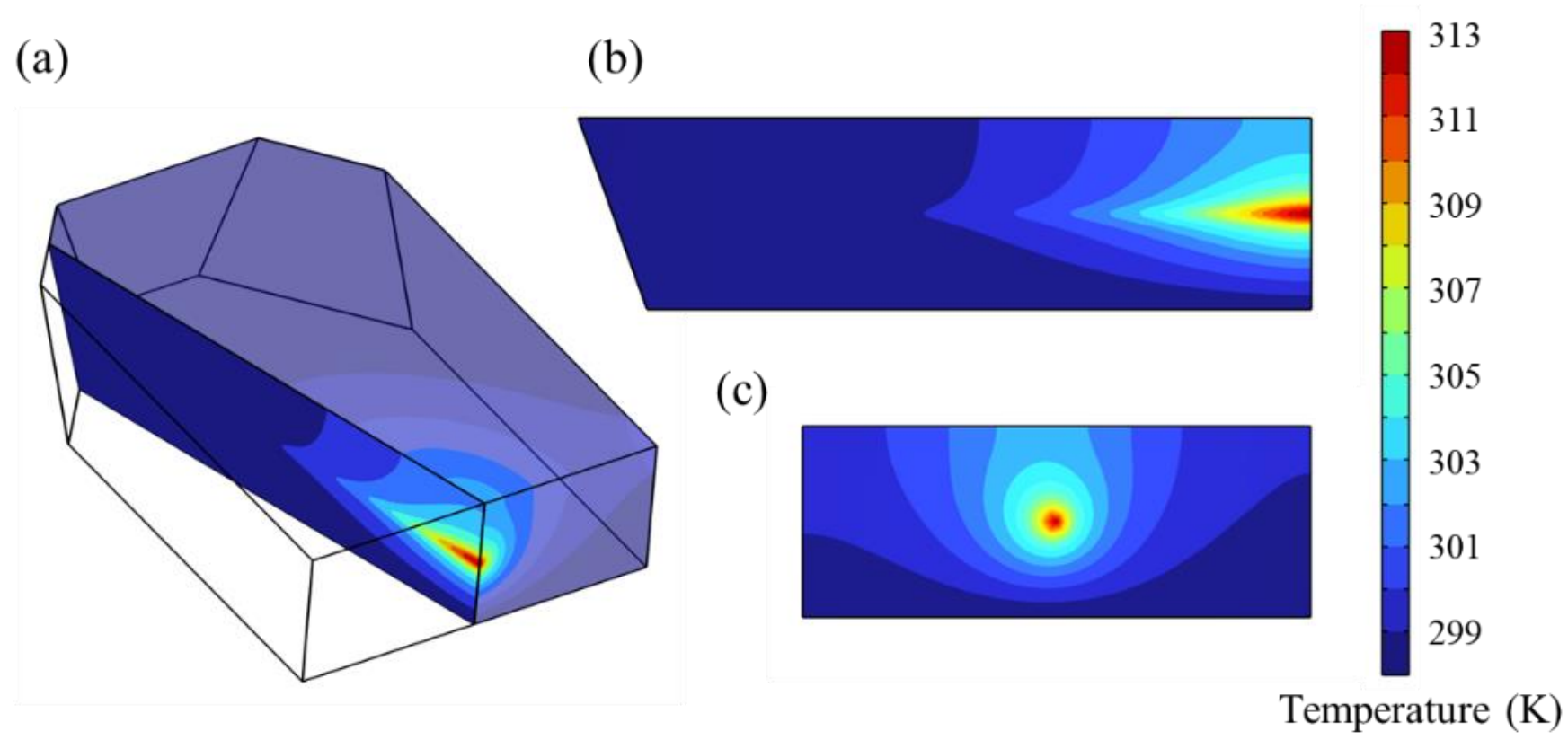

Fig. 7. (a) Temperature distribution within NPRO. (b) Temperature distribution of axial section along leg *AB*. (c) Temperature distribution of *A*-facet. In the simulations, $P_{th}$=100 mW, $\eta_{slope}$ = 60%, $\omega_p$ = 0.1 mm, $\alpha$ = 4.6 cm$^{-1}$. The temperature control of the resonator base is set to 25 °C. Other surfaces dissipate heat through air convection.

Upon transmission through the resonator illustrated in Fig. 7, the beam undergoes focusing. By employing ray-tracing-method as shown in Fig. 8(a), we can calculate the equivalent focal length of thermal-lens induced by the heterogeneous temperature distribution within the resonator. By comparing the Gaussian beam radius (1/e$^2$) of the beam profile at the *A*-facet before (0.2 mm) and after (0.19 mm) the thermal field effect, the equivalent focal length of the thermal-lens at a pump power of 1000 mW is calculated to be approximately 1352 mm, as demonstrated in Fig. 8(b). This is achieved by analyzing the alterations in the spot size and shape of the output beam after it traverses the resonator in one round-trip. This analysis provides critical insights into the optical path modifications induced by thermal effects within the gain medium. By combining thermodynamic simulations with ray-tracing-method, we can calculate the equivalent focal length of the thermal-lens with varying pump power, as shown in Fig. 8(c). One can see the equivalent focal length decrease from 6.35 m to 0.25 m with pump power increased from 0.2 W to 10.2 W.

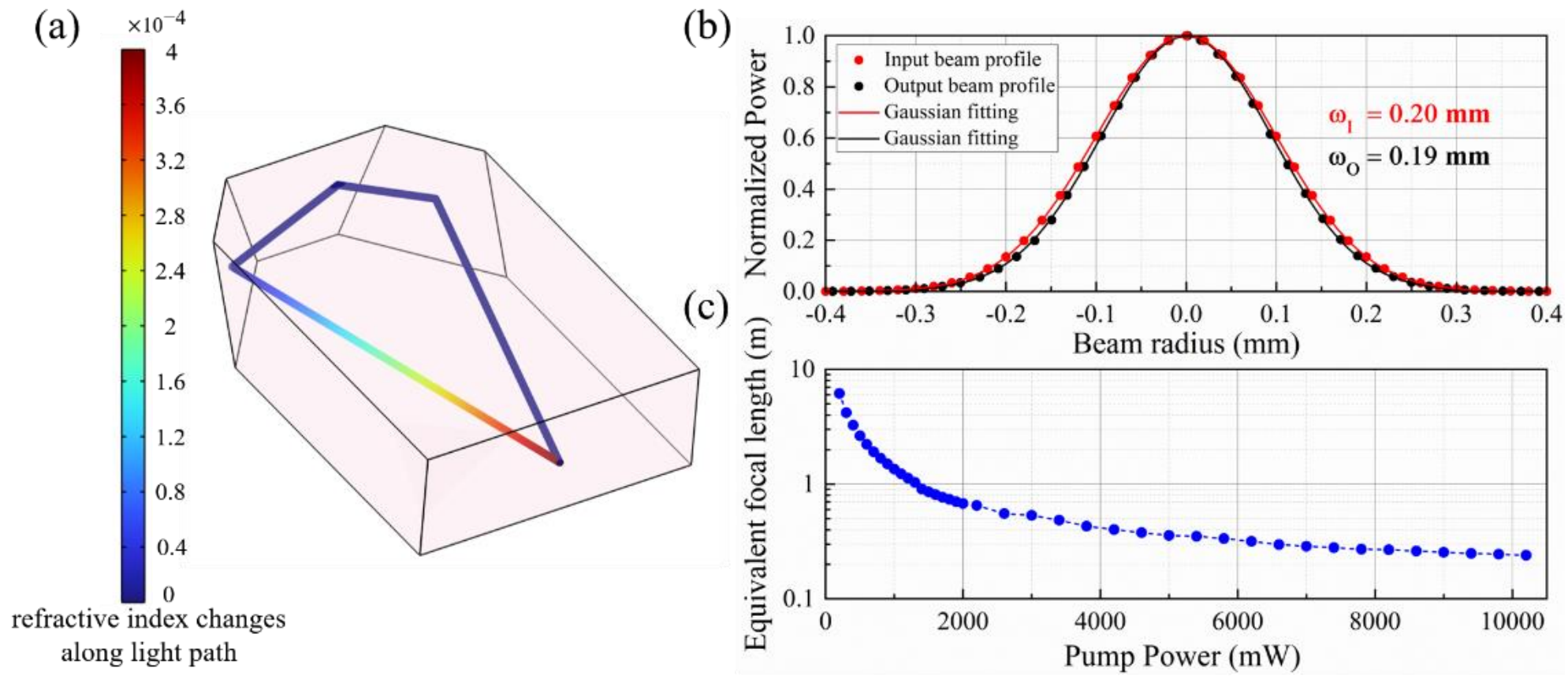

Fig. 8. (a) Equivalent focal length calculation using tracing ray in the resonator. The color bar shows the refractive index changes along the light path under 1000 mW pump power. (b) Normalized power distribution versus beam radius. Red dots and Gaussian fitting: input transmitting beam profile without pump laser heating; Black dots and Gaussian fitting: output transmitting beam profile with pump laser heating. The beam radius $\omega_I$ and $\omega_O$ at $1/e^2$ of the beam power are given by the fittings. (c) Equivalent focal length of the thermal-lenses under different pump power.

We utilized the theoretical model presented in Section 3 to calculate the impact of thermal-lens effects on the output beam profiles, which are shown in Fig. 9. Our simulations indicate that the beam field has good symmetry under pump power of 12.0 W, and the asymmetry arises after the pump powers are beyond 12.0 W, where the $f_A$ is lower than 100 mm. In practical applications, this situation may lead to non-resonance within the laser cavity. For high-power pumped NPRO laser, to mitigate severe thermal-lens effects, gradient doping or using a negative curvature lens at the *A*-facet can be employed. Furthermore, a slighter thermal effect correlates with a larger beam profile on the output surface, which is consistent with general expectations.

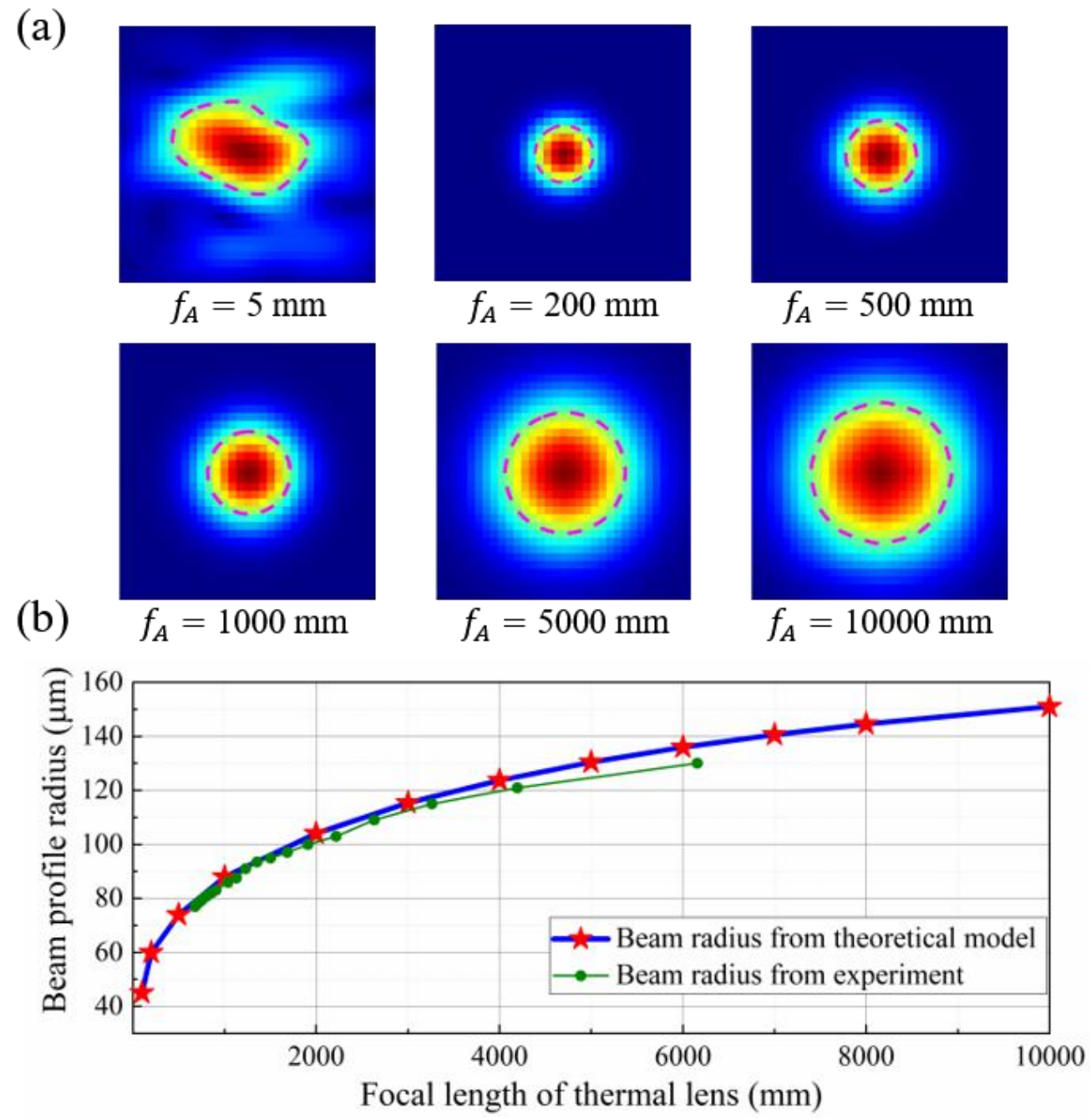

Fig. 9. (a) Beam profiles at point ***S*** with different focal length of thermal-lens. (b) Beam profile radius of different thermal-lenses tested from theoretical model and experiment. In the simulations, $AE$=10.5 mm, $CE$=1.5 mm, $\beta$=90°, $\theta_A$=15°, $f_B$=$10^5$ m.

In experimental measurements, it is not possible to directly measure the beam profile at *A*-facet of NPRO. To validate the simulation results, we measured the beam radius of output laser and its divergence angle from the NPRO, which enables us to resolve the beam radius near *A*-facet. The schematic of the experimental setup is illustrated in Fig. 10. The pump source is a multimode laser diode with fiber-pigtail output, which provides a maximum pump power of 2 W at 808 nm. The pump beam from the laser diode is injected to the Nd:YAG NPRO by a single focus-lens. A pair of neodymium magnets generates sufficient magnetic field for unidirectional emission at 1064 nm [38]. Temperature stabilization for both the laser diode and the NPRO is provided by thermoelectric cooler modules. The output laser beam profile is measured using a CMOS camera (Thorlabs, BC207VIS).

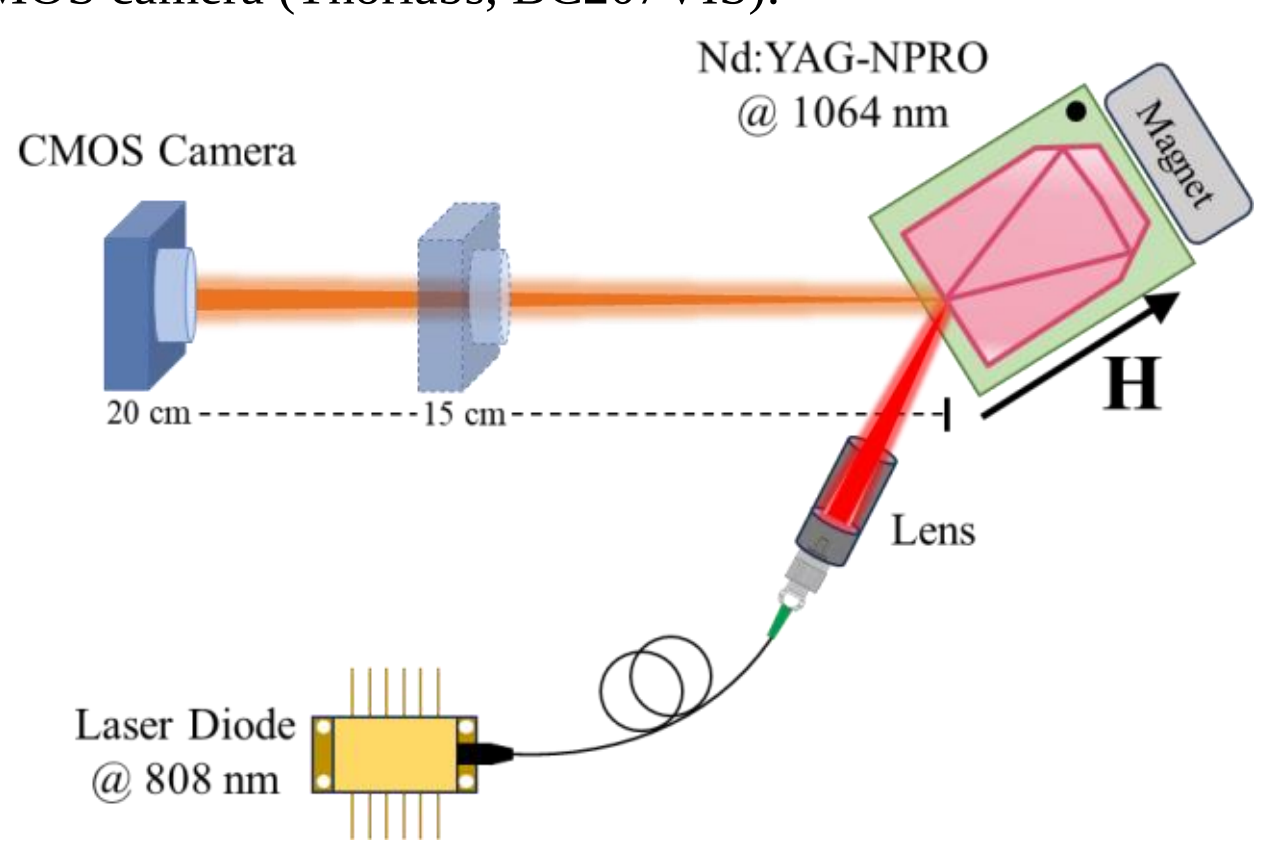

Fig. 10. The schematic of the experimental setup for beam profile measurements.

We varied the pump power and measured the radius of the output beam at distances of 15 cm and 20 cm from *A*-facet of the NPRO. These data are used to approximately calculate the divergence angle of the output beam by Eq. (12). The results are shown in Fig. 11, where the pump power increases from 200 mW to 2000 mW, and the divergence angle of the output laser also increases significantly from 0.16° to 0.40°.

$$\text{Divergence Angle} = \arctan\left(\frac{R_{20\text{ cm}} - R_{15\text{ cm}}}{5\text{ cm}}\right) \tag{12}$$

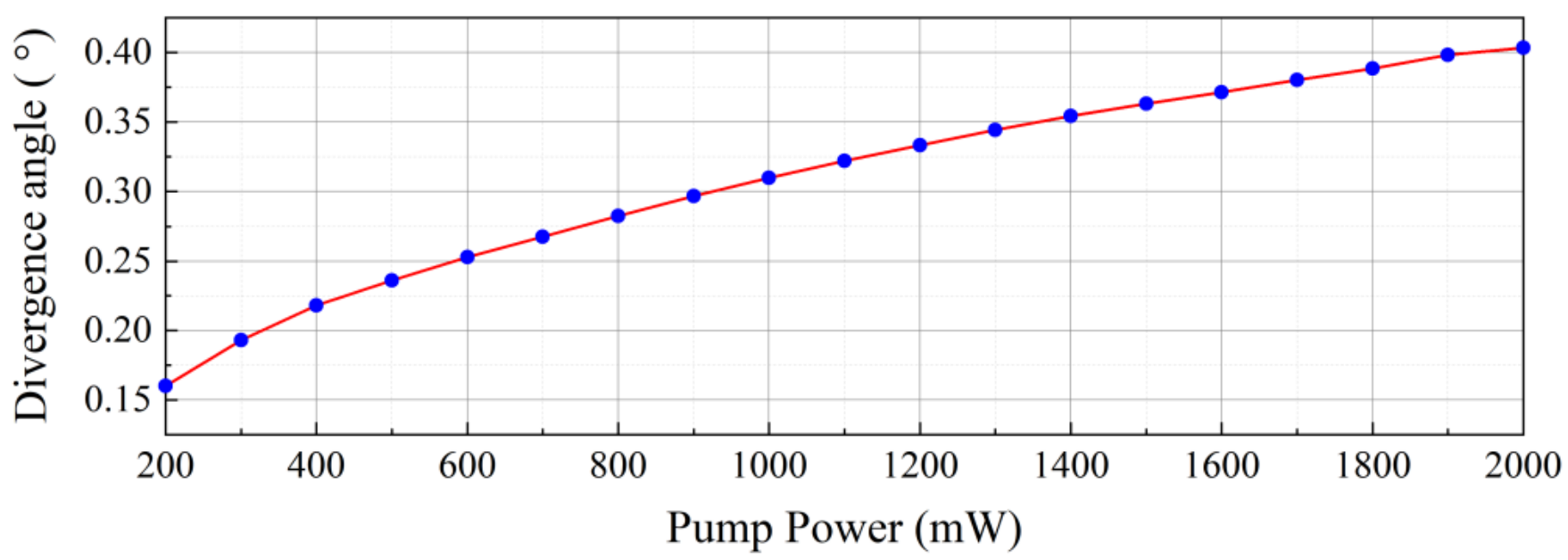

Fig. 11. Divergence angles of the output beam measured under different pump power.

Using the divergence angle of the output laser and the beam radius at the distance of 15 cm or 20 cm, we can calculate the beam radius at the *A*-facet corresponding to different pump powers under the effect of thermal-lens, as shown in Fig. 9(b) by the green dots. They show good match with theoretical simulations with slight discrepancies at large beam profile radius (corresponding to slight thermal-lens effect at low pump power). We attribute this to higher calculation errors of the beam radius under lower pump powers. Although our experiments were limited by the maximum output power of the pump diode, preventing us from measuring results beyond 2 W, the agreement between the available experimental data and simulation calculations allows us to reasonably speculate that the results under high pump-power conditions (> 600 mW) will closely align with those predicted by the simulations. The integration of the output light field model presented in Section 3 with the thermal-lens calculation model in this Section provides valuable guidance for the design of high power NPRO lasers. Although the temperature distribution in the crystal under single-sided cooling exhibits directional non-uniformity, the temperature gradient within the output laser beam radius (radius < 300 μm) is not significant. Consequently, we conclude that the astigmatism introduced by thermal gradients is negligible. This model will be utilized to predict the thermal-lens effects under high pump-power conditions and to mitigate the occurrences that lead to output laser distortions.

## 5. Conclusions

In conclusion, we have developed a transfer matrix for the nonplanar ring laser loop in monolithic NPRO and calculated the output laser beam profile using FEM to discretize the cavity mirrors. Our analysis of various geometrical parameters revealed that the lengths of *AE* and *CE* can control the size of the beam profile, while the laser incidence angle does not alter the symmetry of the beam profile. Additionally, we focused on the effects of thermal-lens within the monolithic NPRO. By combining optical simulations with solid-state heat transfer calculations, we obtained the temperature distribution within the cavity and calculated the equivalent-focal-length of the thermal-lens. Furthermore, we integrated these results with theoretical model of the output beam profile to assess the effects of different thermal-lenses on the beam profiles, which are well examined by experiments. Our comprehensive analysis on the mode patterns and thermal-lens effect will facilitate the design of monolithic NPROs used for single-frequency lasers and laser gyroscopes, and help to improve their performances in various fields, including but not limited to, fundamental research, precision metrology, high-end industry, and deep-space exploration.

**Funding.** National Natural Science Foundation of China (12404489); Fundamental Research Funds for the Central Universities, Sun Yat-sen University (24QNPY162); National Key Research and Development Program of China (2020YFC2200200); Major Projects of Basic and Applied Basic Research in Guangdong Province (2019B030302001).

**Acknowledgments.**

**Disclosures.** The authors declare no competing interests.

**Data availability.** Data underlying the results presented in this paper are not publicly available at this time but may be obtained from the authors upon reasonable request.